\documentclass[aps,prb,twocolumn,superscriptaddress]{revtex4-2}
\usepackage{amsmath,amssymb,graphicx,bm}
\usepackage{hyperref}

\begin{document}
	
	\title{Geometry-Induced Localization and Multifractality in Spiral quasiperiodic chain}
	\author{Hemant Kumar Sharma}
	\affiliation{School of Physical Sciences, National Institute of Science Education and Research, Jatni 752050, India}
	\date{\today}
	\begin{abstract}
	We study a quasiperiodic Aubry–André (AA) lattice arranged along a spiral curve. In this setup, the changing angle of the spiral naturally stretches and compresses the distances between neighboring sites, which in turn modulates the hopping amplitudes.
	The onsite potential itself remains the familiar AA form, but this geometry-induced variation in the hopping dramatically changes how the system behaves—both in its energy spectrum and in how its states localize.Using inverse-participation ratios together with a full multifractal analysis, we find that curvature makes the system localize much more easily, even at relatively small quasiperiodic strengths. It also produces clear windows where the eigenstates become strongly multifractal. This shows that quasiperiodicity and geometry do not act independently—rather, they reinforce one another in shaping the wavefunctions.	Overall, we observe a smooth evolution of the states: from extended, to multifractal, and finally to strongly localized. Our results pave the way for creating tunable quasiperiodic and geometry-driven localization effects in photonic waveguide arrays, ultracold atoms, mechanical metamaterials, and nanoscale platforms.
	\end{abstract}

	\maketitle

	\section{Introduction}
Quasiperiodic systems constitute a central topic in condensed matter physics and wave dynamics, bridging crystalline order and randomness. Since the seminal works on Anderson localization in disordered media \cite{anderson1958absence} and the formulation of quasiperiodic tight-binding models \cite{harper1955single,aubry1980analytic}, it has been understood that quasiperiodicity can generate rich spectral and transport phenomena distinct from both periodic and random systems. These include fractal energy spectra, critical self-dual points, mobility edges, localization transitions, and multifractal eigenstates \cite{sokoloff1985unusual,modugno2009exponential,ganeshan2015nearest}.

Among such models, the Aubry--Andr\'e (AA) chain occupies a paradigmatic role. It exhibits an exact metal--insulator transition driven by the strength of a quasiperiodic onsite potential, together with critical multifractal behavior at the self-dual point \cite{aubry1980analytic,sokoloff1985unusual,satija2013chern}. The AA model and its generalizations have therefore become standard platforms for exploring the interplay between localization, topology, and quasiperiodicity \cite{kraus2012topological,lohse2018exploring,nakajima2016topological}.

Rapid experimental progress over the last decade has enabled direct realizations of quasiperiodic lattices in a wide variety of physical settings. In ultracold atomic gases, bichromatic optical lattices have demonstrated Anderson localization and AA criticality with high controllability \cite{roati2008anderson,schreiber2015observation}. Related phenomena have also been observed in driven and modulated cold-atom systems, including quantized pumping and topological transport \cite{lohse2018exploring,nakajima2016topological}. Photonic waveguide arrays provide a particularly flexible platform, enabling the observation of localization, topological pumping, and quasiperiodic boundary states \cite{lahini2009observing,plotnik2011observation,verbin2013observation,rechtsman2013photonic}. Comprehensive overviews of discrete photonic lattices and light transport in such media are given in \cite{szameit2010discrete}.

Beyond optics and cold atoms, quasiperiodic concepts have been extended to classical wave systems, including acoustic and mechanical metamaterials, where phononic and elastic excitations emulate electronic quasiperiodicity \cite{apigo2019topological,rosa2019edge}.Plasmonic and moir\'e structures provide yet another arena in which geometry and long-range Coulomb interactions give rise to nontrivial collective excitations, including topological and quasiperiodic plasmon modes
\cite{woods2014commensurate,ni2018plasmon,jin2017topological,song2016chiral}.

In parallel, the role of geometry and curvature in quantum and wave systems has attracted growing interest. Curved and strained lattices can generate effective gauge fields, modify hopping amplitudes, and reshape spectral and topological properties, as demonstrated in strained graphene and related Dirac materials where lattice deformations act as synthetic gauge fields \cite{guinea2010energy,vozmediano2010gauge,cortijo2015elasticity}. More generally, geometric and quantum metric effects have been shown to influence transport and topological responses in crystalline systems \cite{can2014geometry}. Such geometric effects are experimentally accessible in several platforms, including photonic lattices and other synthetic systems where lattice geometry and curvature can be precisely engineered \cite{rechtsman2013photonic,ozawa2019topological}.
 In photonics, three-dimensional waveguide architectures enable helical and Floquet lattices with engineered curvature \cite{szameit2010discrete}. In ultracold atoms, optical vortex and Laguerre--Gaussian beams allow the realization of ring and spiral trapping geometries \cite{wright2000optical,ortix2011quantum,bakke2012torsion,ozawa2019topological}.

At the nanoscale, chiral and spiral molecular structures introduce geometry-driven transport phenomena, including chiral-induced spin selectivity and spin-dependent localization \cite{naaman2012chiral,guo2012spin}. These systems highlight how structural chirality can strongly influence electronic and spin dynamics.

Motivated by these developments, we explore how embedding a quasiperiodic Aubry--Andr\'e chain into a spiral geometry modifies its spectral, localization, and multifractal properties. While the onsite quasiperiodic modulation retains the standard AA form, the curved embedding induces systematic variations in intersite distances and hence in the hopping amplitudes, yielding a geometry-controlled inhomogeneous tight-binding model. We demonstrate that curvature alone can significantly enhance localization at reduced quasiperiodic strengths, reshape the energy spectrum, and alter the multifractal characteristics of the eigenstates. Our results establish spiral geometries as a simple and experimentally viable route to engineer quasiperiodicity and to control localization and multifractality across photonic, atomic, mechanical, and nanoscale chiral platforms.

\section{Model and Hamiltonian}

We consider a one-dimensional tight-binding chain embedded in a planar spiral.  
The Hamiltonian is
\begin{equation}
	H = 
	\sum_{i=0}^{L-2} 
	\left( t_i\, c_{i+1}^\dagger c_i + \mathrm{H.c.} \right)
	+ \sum_{i=0}^{L-1} 
	V_i\, c_i^\dagger c_i ,
	\label{eq:H}
\end{equation}
where $t_i$ denotes the nearest-neighbor hopping amplitude and $V_i$ the onsite potential.

\subsection{Spiral embedding}

Each site is placed at
\begin{equation}
	\mathbf{r}_i = 
	a\, i \begin{pmatrix}
		\cos\theta_i \\[2pt]
		\sin\theta_i
	\end{pmatrix},
	\label{eq:ri}
\end{equation}
with angular coordinate
\begin{equation}
	\theta_0 = 0, \qquad 
	\theta_i = n \sum_{k=1}^i \frac{\Delta\theta}{k}.
	\label{eq:theta}
\end{equation}
This nonlinear winding generates a tightly coiled inner region and a gradually opening outer region, giving rise to a broad distribution of bond distances.

The distance between sites $i$ and $i+1$ is
\begin{equation}
	d_i = \lVert \mathbf{r}_{i+1} - \mathbf{r}_i \rVert ,
	\label{eq:di}
\end{equation}
and the geometry-induced hopping amplitude takes the form
\begin{equation}
	t_i = t_0 \exp(-d_i/\xi),
	\label{eq:ti}
\end{equation}
where $t_0$ is the reference hopping and $\xi$ is the decay length.

\subsection{Quasiperiodic modulation}

The onsite potential is the standard Aubry-André modulation, 

\begin{equation}
	V_i = \lambda \cos(2\pi \beta i + \phi),
	\label{eq:Vi}
\end{equation}

with quasiperiodic strength $\lambda$, an irrational parameter $\beta$ (taken as the inverse golden ratio), and phase offset $\phi$. Since $V_i$ remains constant with the embedding, any changes to the spectral and localization properties come solely from the geometry-driven variation in the hopping amplitudes.

The spiral embedding changes the lattice's geometry by allowing the angular coordinate to accumulate nonlinearly along the chain. When the increment parameter $\Delta\theta$ is set to zero, Fig.~\ref{fig:geometry_comparison} shows that the lattice becomes a straight one-dimensional chain with uniform nearest-neighbor separations. This results in a homogeneous hopping profile. However, for nonzero $\Delta\theta$, the angular displacement increases as a harmonic sum, creating a smoothly curved geometry where intersite distances vary systematically with position. Since the onsite AA potential remains unchanged, all geometry-induced effects stem from the corresponding modulation of the hopping amplitudes.

Unlike models where the hopping amplitudes are defined by an arbitrary functional form, the modulation in the spiral-embedded Aubry-André lattice is fully determined by geometry through the relation \(		t_i = t_0 \, e^{-d_i/\xi}.\) Thus, the hopping profile is not an independent control parameter; instead, it results from the nonlinear accumulation of the angular coordinate and the resulting site-to-site distances. This geometric constraint creates correlated variations among neighboring hopping amplitudes that cannot be replicated by simply applying an off-diagonal modulation. Specifically, the harmonic structure of the spiral ensures that the spatial variation of $t_i$ is smooth, monotonic, and guided by curvature, rather than freely adjustable or random. As a result, the geometry reshapes both the effective kinetic-energy scale and the relative strength of the quasiperiodic potential in a self-consistent way. This creates a localization landscape that is fundamentally different from standard hopping-modulated AA models, where onsite and off-diagonal terms can be adjusted independently. In contrast, the spiral embedding offers a physically motivated and experimentally accessible mechanism that enforces a correlated hopping modulation, leading to curvature-assisted localization and multifractality that cannot be achieved from typical inhomogeneous hopping profiles.
	
	\begin{figure}[t]
		\centering
		\includegraphics[width=0.85\linewidth]{spiral_two.png}
		\caption{
			Geometry of the spiral-embedded AA lattice for two values of the angular
			increment parameter $\Delta\theta$.  
			Top: for $\Delta\theta_1 = 0$, the chain is straight and all nearest-neighbor 
			separations are equal, giving rise to uniform hopping amplitudes.  
			Bottom: for $\Delta\theta_2 = \pi/20$, the accumulated angular rotation leads 
			to smoothly varying bond lengths and a corresponding modulation of the hopping.  
		}
		\label{fig:geometry_comparison}
	\end{figure}
	\begin{figure}[t]
		\centering
		\includegraphics[width=\linewidth]{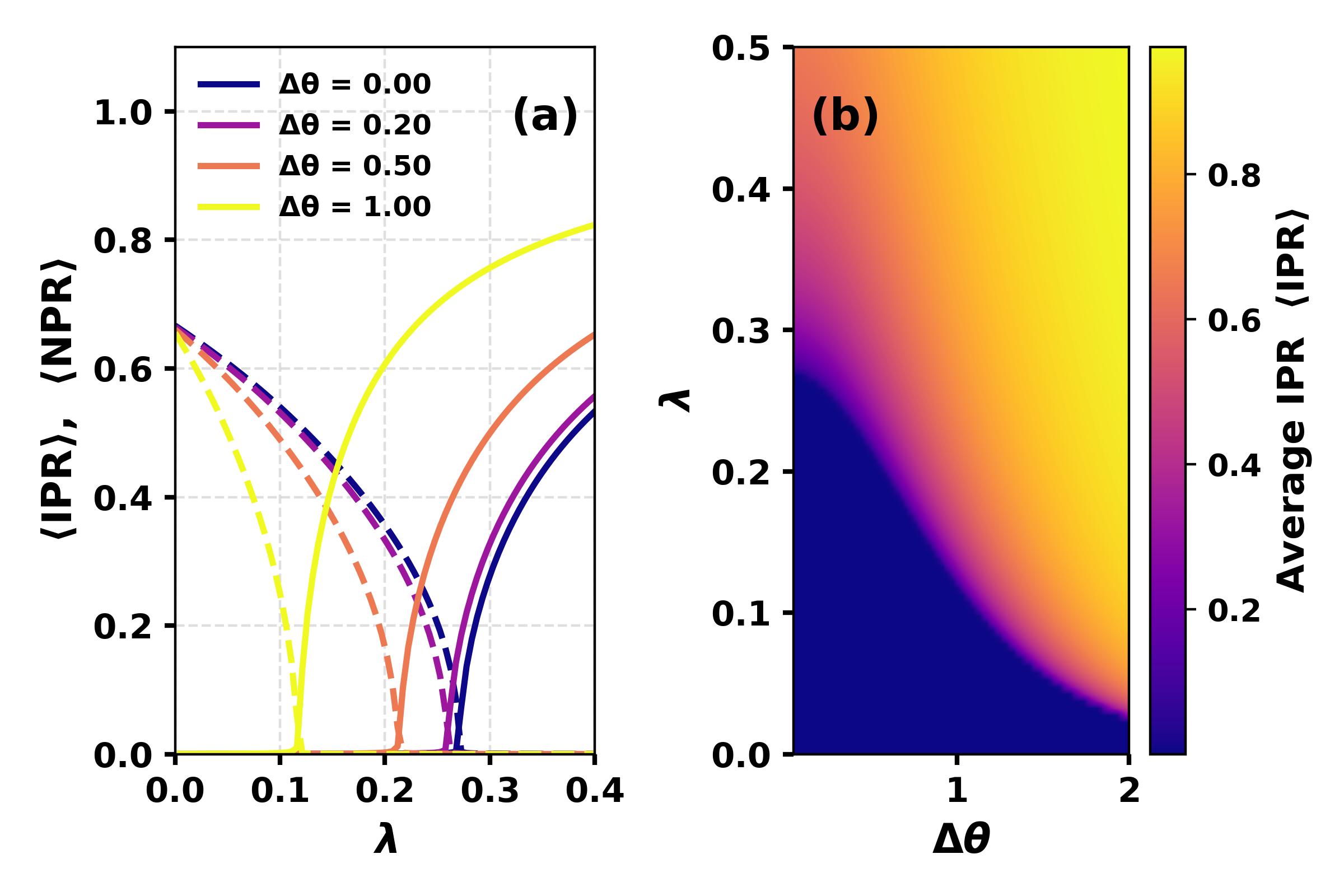}
		\caption{
			(a) Average inverse participation ratio (IPR, solid lines) and normalized 
			participation ratio (NPR, dashed lines) as functions of the quasiperiodic 
			strength $\lambda$ for several values of the angular increment parameter 
			$\Delta\theta$.  
			(b) Phase diagram of the average IPR in the $(\lambda,\Delta\theta)$ plane. Here $t_0 =  1$,$\xi= 0.5$, $n= a=1$}
		\label{fig:IPR_phase_diagram}
	\end{figure}
Figure~\ref{fig:IPR_phase_diagram} shows how geometric deformation changes the localization behavior of the spiral-embedded AA lattice. In this setup, the parameter $\Delta\theta$ controls how quickly the angular coordinate accumulates along the chain. This parameter affects the curvature and the geometry-induced changes in the hopping amplitudes. When $\Delta\theta$ is small, the spiral has weak curvature, the nearest-neighbor distances remain almost uniform, and the model acts similarly to the conventional AA chain with a largely uniform hopping profile. In this situation, the localization transition happens at a relatively high value of the quasiperiodic strength $\lambda$, approaching the well-known criterion of the standard AA model, $\lambda_{\mathrm{c}} = 2t$, where $t$ is a constant hopping amplitude.

As $\Delta\theta$ increases, the nonlinear angular accumulation leads to more nonuniform bond lengths. Since the hopping amplitudes decrease exponentially with distance, the resulting hopping profile becomes very inhomogeneous. This effectively reduces the average hopping and introduces large local variations in the kinetic energy scale. Because the critical point of the AA model is sensitive to the ratio $\lambda / t$, any decrease in the effective hopping strength $t_{\mathrm{eff}}$ due to geometric distortion shifts the localization threshold to lower values of $\lambda$. In other words, the well-known condition $\lambda_{\mathrm{c}} = 2t$ changes to a geometry-modified criterion,

\[
\lambda_{\mathrm{c}} \sim 2 t_{\mathrm{eff}}(\Delta\theta),
\]

where $t_{\mathrm{eff}}(\Delta\theta)$ decreases steadily as the spiral curvature increases. This explains why even small values of $\Delta\theta$ can cause localization at values of $\lambda$ much lower than the transition point of the straight chain.

For sufficiently large $\Delta\theta$, the curvature-induced changes in the hopping dominate over the quasiperiodic onsite potential. The system becomes localized across almost the entire range of $\lambda$ shown in the figure. Thus, curvature and quasiperiodicity work together: the quasiperiodic potential creates an aperiodic onsite landscape, while the spiral geometry effectively increases this disorder by changing the strength of quantum tunneling. This combination allows geometry alone to push the system into a localized phase, even when the intrinsic quasiperiodicity is weak, showing that the spiral embedding offers a flexible way to control localization.
\begin{figure}[t]
	\centering
	\includegraphics[width=\linewidth]{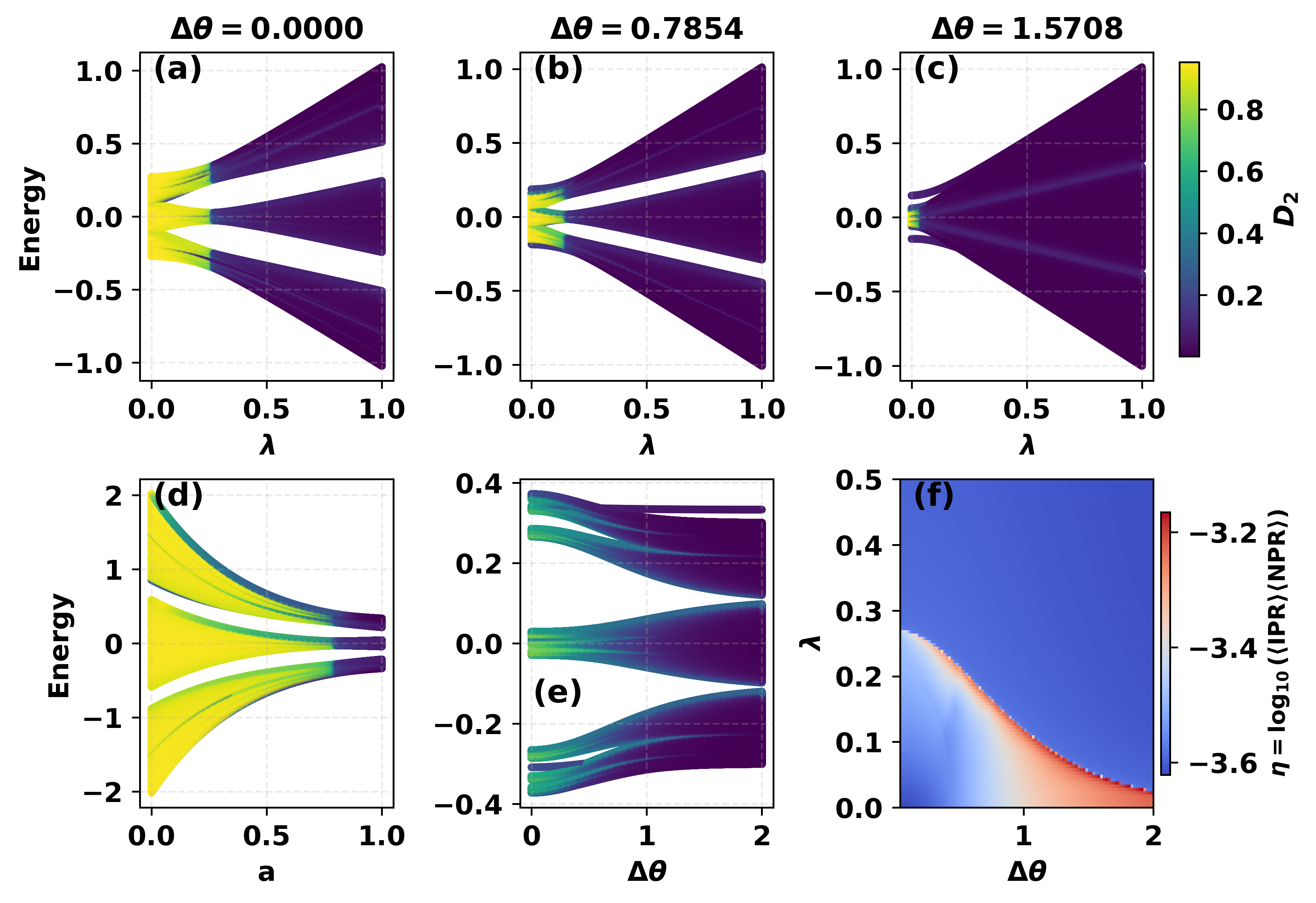}
	\caption{
		Energy spectra and multifractal characteristics of the spiral-embedded Aubry--André model for varying quasiperiodic strength $\lambda$, angular increment $\Delta\theta$, and radial scale $a$. The colors represent either the generalized fractal dimension $D_2$ or the phase indicator $\eta$, highlighting the evolution from extended to multifractal to localized states under geometric deformation.
	}
	
	\label{fig:g2_fractaldimension}
\end{figure}
Figure~\ref{fig:g2_fractaldimension} illustrates how geometric deformation profoundly modifies the multifractal and localization characteristics of the spiral-embedded Aubry--André model. Panels (a)–(c) show the energy spectrum as a function of the quasiperiodic strength $\lambda$ for different angular increments $\Delta\theta$, with the color scale representing the generalized fractal dimension $D_2$. For $\Delta\theta=0$, the model reduces to the standard AA chain, where all states are extended for $\lambda<1$, localized for $\lambda>1$, and display multifractal behaviour only at the critical self-dual point $\lambda=1$. This sharp transition is reflected by the narrow region of intermediate $D_2$ values in panel (a). However, once the system is embedded on a spiral, the situation changes qualitatively: increasing $\Delta\theta$ introduces position-dependent hopping amplitudes, which break the exact self-duality of the AA model. As a result, the single critical point broadens into a finite parameter window in which states acquire multifractal properties. This is seen in panels (b) and (c), where $D_2$ no longer peaks only at a single $\lambda$ but extends over an enlarged region of the spectrum. At the same time, the geometric stretching reduces the effective hopping amplitude, causing the overall magnitude of $D_2$ to decrease. By the time $\Delta\theta=\pi/2$, multifractality is almost entirely suppressed and the spectrum is dominated by localized states even for comparatively weak quasiperiodic potentials, demonstrating that curvature strongly enhances localization.

Panel (d) shows the complementary role of the radial growth parameter $a$. Increasing $a$ introduces stronger spatial inhomogeneity in the intersite distances, leading to a more rapidly varying hopping profile. This further weakens the effective kinetic energy scale and suppresses $D_2$ across all energies, pushing the system toward localization even when $\lambda$ is held fixed. Panel (e) isolates the pure geometric contribution by plotting the spectrum against $\Delta\theta$ at a fixed $\lambda$. Here, one clearly observes a continuous decay of $D_2$ as curvature increases: states that are partly multifractal at small $\Delta\theta$ evolve smoothly into fully localized states as geometric deformation intensifies, confirming that curvature alone can drive the multifractal-to-localized transition independently of the quasiperiodic strength.

Finally, panel (f) presents a global view of the interplay between quasiperiodicity and geometry via the phase diagram in the $(\Delta\theta,\lambda)$ plane, constructed using the indicator $\eta=\log_{10}(\langle\mathrm{IPR}\rangle/\langle\mathrm{NPR}\rangle)$. This measure cleanly distinguishes extended, multifractal, and localized phases. The resulting mobility boundary slopes downward as $\Delta\theta$ increases, showing that geometric deformation reduces the effective hopping scale and shifts the localization transition to significantly smaller values of $\lambda$ than in the conventional AA model. Thus, the spiral embedding not only broadens the critical point into an extended multifractal region but also greatly enhances localization through purely geometric effects.

	\begin{figure}[t]
		\centering
		\includegraphics[width=\linewidth]{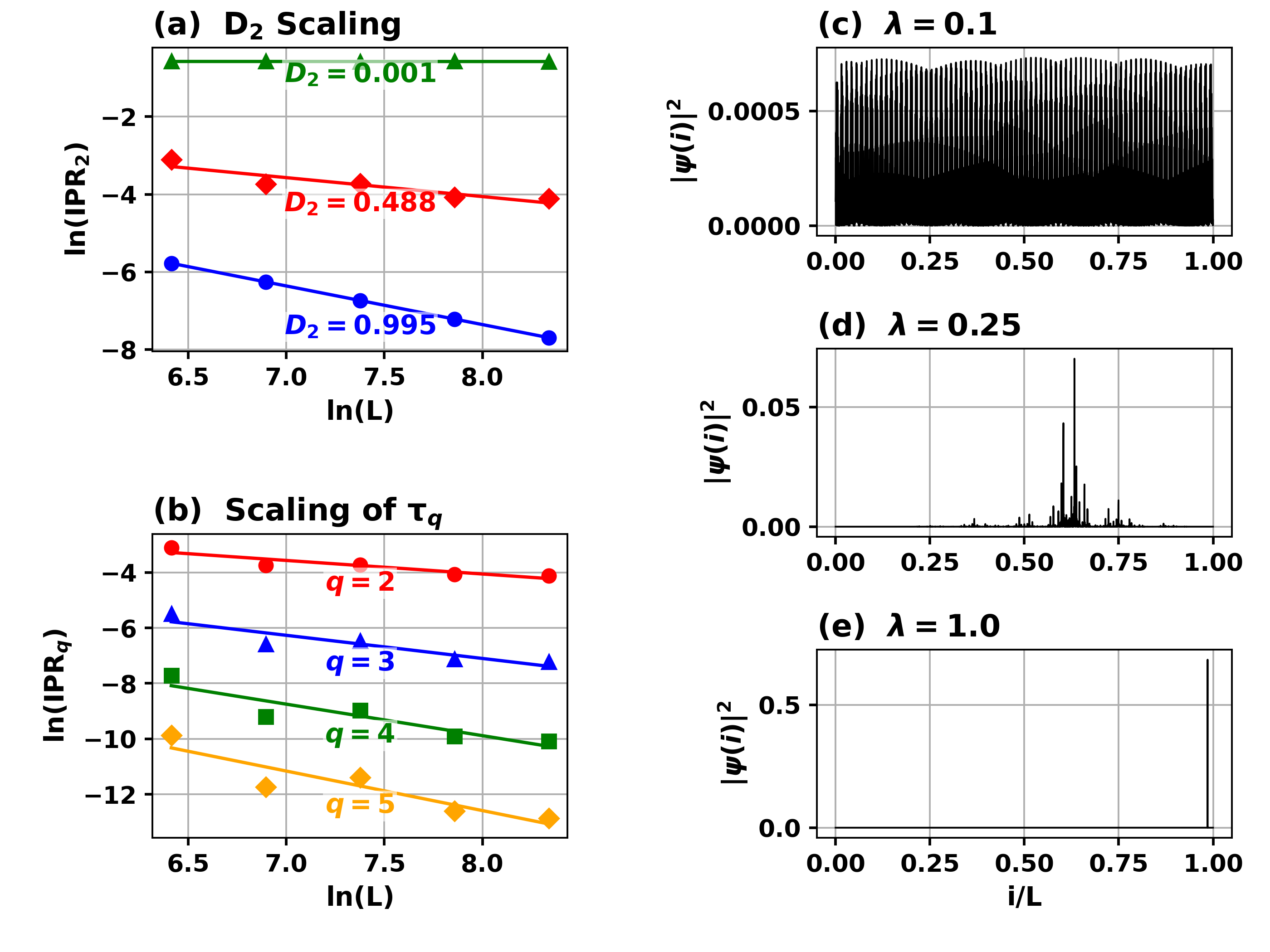}
		\caption{
			Multifractal characterization of eigenstates in the spiral-embedded AA lattice.
			(a)  $\ln(\mathrm{IPR}_2)$ with $\ln(L)$ for representative eigenstates.
			The slopes yield the fractal dimension $D_2$: the blue line ($D_2 \approx 0.995$)
			corresponds to an extended state, the red line ($D_2 \approx 0.488$) to a
			multifractal state, and the green line ($D_2 \approx 0.001$) to a localized
			state.  
			(b)  $\ln(\mathrm{IPR}_q)$ for $q=2,3,4,5$.    
			(c)--(e)  $|\psi(i)|^2$ with increasing quasiperiodic
			strength $\lambda$.  
			For $\lambda = 0.1$ (c), the wavefunction spreads across the entire system with
			small fluctuations, consistent with $D_2 \approx 1$.  
			At $\lambda = 0.25$ (d) with intermediate 
			$D_2$.  
			For $\lambda = 1.0$ (e), $D_2 \approx 0$. Here $\delta_\theta = 0.285$,$t_0 =  1$,$\xi= 0.5$, $n= a=1$ }
		\label{fig:multifractal_results}
	\end{figure}

	To describe the localization behavior of the eigenstates in a way that goes beyond standard inverse participation ratio (IPR) measures, we use a detailed multifractal analysis. This approach is essential for understanding critical states in quasiperiodic and engineered lattices \cite{lahini2009observing,roati2008anderson,kraus2012topological}. 
	
	For an eigenstate $\psi_i^{(m)}$, the generalized inverse participation ratio is defined as
	\begin{equation}
		\mathrm{IPR}_q^{(m)} = \sum_{i=1}^{L} |\psi_i^{(m)}|^{2q} 
		\propto L^{-\tau_q},
	\end{equation}
	where $q$ is a real number and $\tau_q$ is the corresponding mass exponent. Extended states satisfy $\tau_q = d(q-1)$ with $d=1$, which aligns with the ballistic spreading observed in ultracold-atom and photonic experiments on quasiperiodic lattices \cite{roati2008anderson,schreiber2015observation}. Localized states, on the other hand, have $\tau_q=0$. Multifractal states, known to emerge at the localization transition of the Aubry–André model and seen directly in photonic waveguide measurements \cite{lahini2009observing}, show a nonlinear dependence $\tau_q = D_q(q-1)$, where $D_q$ represents the generalized fractal dimension. Here, $D_q=1$ indicates an extended state, $D_q=0$ indicates a localized state, and values in between show a nonuniform fractal distribution of amplitudes. 
	
	In our analysis, we focus on the correlation (fractal) dimension $D_{2}$, which we get from the scaling law $\mathrm{IPR}_{2} \propto L^{-D_{2}}$. This serves as a direct and reliable way to distinguish between extended, multifractal, and localized eigenstates. Figure~\ref{fig:multifractal_results} illustrates this behavior across different quasiperiodic strengths. 
	
	Panel~(a) shows how $\ln(\mathrm{IPR}_{2})$ scales with $\ln(L)$ for three representative eigenstates. A slope near one corresponds to $D_{2} \approx 1$, indicating an extended state with amplitude spreading uniformly across the entire lattice. Such states have been observed in ultracold-atom and photonic realizations of quasiperiodic systems \cite{roati2008anderson,lahini2009observing}. A slope between zero and one yields $0 < D_{2} < 1$, showing a multifractal wavefunction where the amplitude is concentrated on a sparse yet significant set of lattice sites. This behavior marks criticality in quasiperiodic lattices and has been experimentally detected in laser-written photonic quasicrystals \cite{amaral2018synthetic,verbin2013observation}. Finally, a nearly flat line indicates $D_{2} \approx 0$, typical of a fully localized state confined to just a few sites. 
	
	Panel~(b) expands on this characterization by showing the generalized scaling $\ln(\mathrm{IPR}_{q})$ for $q=2$ to $5$. The fact that the extracted mass exponents $\tau_{q}$ do not fall on a straight line as a function of $q$ shows real multifractal behavior rather than a simple shift between extended and localized phases. This nonlinear $\tau_{q}$–$q$ dependence is a hallmark of intermediate critical states, consistent with both theoretical predictions and experimental findings in quasiperiodically modulated photonic lattices and synthetic matter platforms. 
	
	Panels~(c) to (e) display the corresponding real-space probability densities $|\psi(i)|^{2}$, showing how the wavefunctions change with increasing quasiperiodic strength. For $\lambda=0.1$ [panel~(c)], the probability density is almost uniformly distributed across the entire system. This confirms the extended nature of the eigenstate. For $\lambda=0.25$ [panel~(d)], the wavefunction shows strong spatial fluctuations and distinct peaks, indicating a multifractal structure with complex self-similar scaling. For $\lambda=1.0$ [panel~(e)], the wavefunction concentrates in a very small part of the lattice. This matches the extracted value $D_{2} \approx 0$ from the scaling analysis and indicates a fully localized phase. 
	
	Together, these observations reveal a clear shift from extended to multifractal to strongly localized eigenstates as the quasiperiodic modulation increases. Importantly, this behavior occurs within a spiral geometry, where the effective hopping inhomogeneity is created solely by geometric embedding. Because these geometric modulations can be made in laser-written waveguide arrays \cite{szameit2010discrete} and curved photonic lattices \cite{rechtsman2013photonic}, our results show a realistic and experimentally feasible way to create tunable multifractality and geometry-driven localization.
	
	\section{Conclusion}
	
In this work, we investigated how embedding a quasiperiodic Aubry--André lattice along a spiral curve alters its spectral, localization, and multifractal properties. The nonlinear evolution of the angular coordinate generates a systematic variation in the distances between neighboring sites, which in turn produces strongly position-dependent hopping amplitudes while leaving the onsite AA potential unchanged. This geometric inhomogeneity transforms the model from a uniform tight-binding chain into a spatially modulated one, and our analysis shows that this curvature-induced modulation significantly enhances localization. In particular, the onset of the localized phase shifts to values of the quasiperiodic strength far below the standard self-dual point of the simple AA model. The spiral geometry also reshapes the energy spectrum, producing broadened bands and geometric distortions that have no analogue in straight chains.

A central result of our study is that curvature qualitatively modifies the multifractal properties of the eigenstates. In the simple AA model, multifractality appears only at the self-dual critical point. In contrast, the spiral embedding breaks exact self-duality by introducing spatially varying hopping, thereby broadening the single AA critical point into a finite parameter region that supports multifractal states. Our scaling analysis of the generalized inverse participation ratios captures this transition clearly: as the quasiperiodic strength increases, the system evolves smoothly from extended to multifractal and eventually to localized states. The correlation dimension $D_{2}$ highlights this extended critical window, while the nonlinear mass exponents $\tau_{q}$ confirm that the observed multifractality is a genuine geometry-assisted effect. These results demonstrate that curvature can act as an independent control knob for tuning fractal characteristics of quantum states.

Overall, our findings reveal that spiral embeddings offer a simple and versatile geometric mechanism for engineering localization and multifractality in quasiperiodic systems. Because spiral and helical geometries can be readily implemented in experimental platforms such as laser-written photonic waveguides, optical vortex traps, twisted atomic lattices, and chiral nanoscale structures, our results provide a realistic route toward exploring geometry-driven criticality, transport, and wave control. Future directions include the study of interactions in curved quasiperiodic lattices, the behavior of topological boundary modes in spiral geometries, and extensions to higher-dimensional or dynamically modulated curved lattices.

\end{document}